\documentclass[12pt]{article}

\usepackage{natbib}
\usepackage{amsmath}
\usepackage{amsfonts}
\usepackage{amssymb}
\usepackage{latexsym}
\usepackage{a4}
\usepackage{psfrag}
\usepackage{ulem}
\usepackage{epsfig}

\def\P{{\text P}}
\def\T{{\text T}}
\def \ii{{\mathrm{i}}}
\def\e{{\text e}}

\def \D{{\mathrm{D}}}

\def \LL{{\cal{L}}}

\def \RR{{\cal{R}}}

\def \pd{\partial}

\def \UU{\boldsymbol{U}}

\def \Bbeta{{\boldsymbol{\beta}}}

\def \Bx{{\boldsymbol{x}}}
\def \Bv{{\boldsymbol{v}}}
\def \Bu{{\boldsymbol{u}}}

\begin{document}


\title{{\bf The elastodynamic model of wave-telegraph type for quasicrystals}}

\author{Eleni Agiasofitou\footnote{{\it E-mail address:} agiasofitou@mechanik.tu-darmstadt.de}\,,\
 Markus Lazar\footnote{{\it E-mail address:} lazar@fkp.tu-darmstadt.de}\\
\\Heisenberg Research Group, Department of Physics, \\Darmstadt University of Technology, \\Hochschulstr. 6, D-64289 Darmstadt, Germany
}

\date{\today}
\maketitle

\begin{abstract}
 \noindent
In this work we propose {\it the elastodynamic model of
wave-telegraph type} for the description of dynamics of quasicrystals. Phonons are represented by waves, and phasons by waves damped in time and propagating with finite velocity. Therefore, the equations of motion
for the phonon fields are of wave
type and for the phason fields are of
telegraph type. The proposed model constitutes a unified theory in the sense that already established models in the literature can be recovered as asymptotic cases of it. The equations of motion for the displacement fields in the theory of incompatible and compatible elastodynamics of wave-telegraph type of quasicrystals are derived.\\


 \end{abstract} 

\section{Introduction}
Quasicrystals were discovered by \citet{Shechtman1984} opening up a new  and promising,
interdisciplinary research field. A comprehensive presentation of the current state of the art
 focusing on the mathematical theory of elasticity can be found in the book of \citet{Fan Book}.
 Within the scientific community of quasicrystals it is well-known that
there has not yet been an agreement, concerning the description of the
dynamic behavior of quasicrystals. Different points of view date
back to \citet{Bak1985a, Bak1985b} and to \citet{Lubensky85}. \citet{Bak1985a, Bak1985b}
argued that phason modes represent structural disorder or
structural fluctuations. \citet{Ding1993} generalized first the classical theory of linear elastodynamics towards quasicrystals. They extended the Newton's law of motion to quasicrystals describing both, phonons and phasons, with wave
propagations. This model can be found in the literature \citep{Fan and Mai, Shi2005, Fan Book} as
{\it elastodynamics of quasicrystals}. To avoid confusion, we propose to use for this model the name {\it elastodynamics of wave type}. On the other hand in analogy to hydrodynamic theory, \citep{Lubensky85, Lubensky86} proposed {\it the hydrodynamics of quasicrystals}
where phonons
and phasons play very different roles in the dynamics, since phasons are insensitive to spatial translations. In general, phasons are diffusive modes rather than propagating. \citet{Rochal2002} proposed a combination of
the two previous models, the so-called {\it minimal model of
the phonon-phason elastodynamics}, according to which the
equations of motion for the phonons are of wave type and for the
phasons are of diffusion type. During the last few years, this
model has been adopted by many researchers (e.g. \citet{RadiMariano2011, Fanetal2012}) and can usually be found under the name {\it
elasto-hydrodynamics of quasicrystals} \citep{Fan2009}.
 \par
 The aforementioned models investigate the phasons in a very different manner. Pure diffusion is very different from wave propagation without attenuation. Our primary motivation has been to formulate a model, which unifies the two approaches (the elastodynamic model of wave type and the elasto-hydrodynamic model) as asymptotic cases of it and moreover provides a theory for the phason dynamics valid in the whole range of possible wavelengths.
\par
We introduce herein {\it the elastodynamic model of wave-telegraph type} for the description of dynamics of quasicrystals. Phonons are represented by waves, and phasons by waves damped in time and propagating with finite velocity; that means the equations of motion for the phonons are partial differential equations of wave type, and for the phasons partial differential equations of telegraph type. In the telegraph type equation we encounter both the first and the second time derivative of the  phason displacement field. In fact the phason velocity field is multiplied by a crucial parameter, which is generally a tensor, the so-called {\it tensor of phason friction coefficients} which is representative for the damping character of phason dynamics and is directly related to the characteristic time of damping of the process. It is interesting that when the phason friction coefficient becomes very large (in comparison with the average mass density of the material per unit time) the telegraph type equation degenerates to a diffusion type one, while when this coefficient becomes small enough, then the telegraph type equation gives birth to the wave type phason dynamics. In other words, the elasto-hydrodynamic model and the elastodynamic model of wave type can be recovered as asymptotic cases of the elastodynamic model of wave-telegraph type by letting the phason friction coefficient to obtain extreme values. This advantage of the telegraph equation has been exploited herein to formulate our approach in the form of a model unifying the existing models and simultaneously offering the possibility to describe phason dynamics for all intermediate values of the phason friction coefficient.
\par
Based on a proper mathematical formulation, that is a variational principal formulation allowing for dissipative processes for the phason fields, we derive the equations of motion for the displacement fields in the theory of incompatible as well as of compatible
 elastodynamics of wave-telegraph type of quasicrystals. The incompatible case is important, for instance, for the
 study of dislocations and plasticity while
the compatible case is important for the study of dynamics of cracks, elastic waves, body forces, as well
as dispersion relations in quasicrystals.

\section{Phason dynamics}
This section is devoted to the description of phason dynamics. We propose that phasons should be represented by waves damped in time and propagating with finite velocity, which corresponds to a telegraph type partial differential equation.
\par
It has been written by \citet{Francoual2003, Francoual2006} in an experimental study of the dynamics of long-wavelength
phason fluctuations in quasicrystals (i-AlPdMn), that phasons are collective diffusive modes. \citet{Francoual2003, Francoual2006}  found that phason waves should
decay exponentially
 in time, with ``characteristic time" (relaxation time) proportional to the square of the phason
 wavelength. In their approach which is based on the hydrodynamics, they use the
 dispersion relation of the diffusion equation producing actually damped standing waves (see eq.~(\ref{standing})) and finding the relaxation time $\tau$ (see eq.~(\ref{theta})). However, for a standing wave,
 there is no oscillation of the examined field, no propagation and
 every term dies away, the terms of shorter wavelength dying away most rapidly \citep{Webster}. Moreover, as it has been well written in \citet{deBoissieu2008} and
\citet{de Boissieu et al 2008} ``hydrodynamics deals with excitations in the long-wavelength limit"  and
``phason modes as a result of the hydrodynamic theory are always diffusive modes, provided the wavelength is large enough".
The question that comes up at this point is what happens in the full
regime of all possible wavelengths and what has to be implemented in order for
the model to incorporate the realistic case of a finite propagation velocity.

\par
The elastodynamic model of wave type can describe phason waves with finite velocity but without attenuation. On the other hand, the hydrodynamic and the elasto-hydrodynamic model support a diffusion type equation for the dynamics of the phason fields. However, as it is well-known the diffusion equation\footnote{Diffusion equation: $\frac{\partial \psi}{\partial t}=\D_{\text {dif}}\, \Delta \psi$,
where $\Delta$ is the Laplacian and $\D_{\text {dif}}>0$ the diffusion coefficient.} supports exclusively an infinite velocity of propagation. For example, when an initial disturbance
 with compact support is applied in space then the signal is transmitted spontaneously with infinite velocity in the whole space.  In fact, using the dispersion relation, it is possible to construct standing waves damped exponentially in time or propagating waves damped exponentially in space but the initial
 disturbance (in the second case) expands at once with infinite velocity to the whole spatial structure of the problem \citep{Kneubuhl}.
 In order to model waves damped exponentially in time and propagating with finite velocity, a telegraph type equation is needed.
 A typical form of {\it the telegraph equation} is given by \citep{MF, Iwanenko, Barton, Kneubuhl}
 \begin{align}    \label{tel}
 \frac{\partial^2 \psi}{\partial t^2} +\frac{2}{\tau_ {\text {tel}}} \frac{\partial\psi}{\partial t}=c^2\Delta \psi, \quad  {\tau_ {\text {tel}}}>0
 \end{align}
 where $\Delta$ is {\it the Laplacian}, $c$ indicates a {\it wave velocity} and $\tau_{\text {tel}}$ a {\it characteristic time of damping}.  The quantity
 $\bf{{\tau_ {\text {tel}}}}$ is also called {\it the modulus of decay of the wave}. Eq.~(\ref{tel}) describes the situation where a wave propagated by the wave equation is subjected to a dissipative effect caused by the term
 $({2}/{\tau_ {\text {tel}}}){\partial\psi}/{\partial t}$ \citep{Jeffrey2003}.
\par
For the description of several physical phenomena, for example wave propagation in cables and wires, heat waves, and oceanic diffusion,
the diffusion equation has been widely used as a first approximation.
However, the diffusion equation, which is a parabolic partial differential equation,
possesses the unphysical property and disadvantage
that admits infinite velocity of propagation \citep{Strauss}. For instance,
the infinite propagation of a thermal signal; that means that the mechanism of heat conduction is
established instantaneously under all conditions, which is a paradox. A proper improvement is obtained by using the telegraph equation,
which is the simplest mathematical model combining waves and diffusion, and it has been proven that it
can describe the real physics more appropriately.  From the mathematical point of view, the telegraph equation
is a hyperbolic partial differential equation overcoming, in this way, the physically inadmissible infinite
velocity of ``information'' propagation. It is worth noting that the first model in the theory of electric signalling, that is for the description of wave
propagation in telegraphs, telephones and cables, in the 19th century was based on the diffusion equation.
It was \citet{Heaviside1876}
(see also~\citet{Heaviside, Griffiths}) who
first understood that this leads to unphysical results and then he proposed his famous equation, the so-called {\it telegraph equation} or
{\it telegrapher's equation}, to improve the theory of telegraphy. For that reason, it was written by Whittaker in the preface of \cite{Heaviside}, that it should be also properly called {\it Heaviside's equation}. Later,
the same ``story'' has been repeated in problems with heat waves~\citep{MF, Vernotte, Cattaneo, Joseph} and
oceanic diffusion~\citep{Okubo1971}.
In the case of heat waves, the telegraph equation is usually called {\it hyperbolic heat conduction equation} (HHC). Sometimes, the name {\it hyperbolic diffusion equation} is also used for the telegraph equation.
\par
For the modeling of damped phason waves the concept of a phason frictional force is adopted. This force is given in terms of an anisotropic tensor, the tensor of the phason friction coefficients, in order to capture anisotropic effects of the damped waves. The anisotropic structure of this tensor is inspired by \citet{Francoual2003}, since in experiments strong anisotropies in the phason diffuse scattering intensity have been observed. Therefore, the proposed model accounts also for anisotropy; in the sense that the phason waves can be damped anisotropically.


\section{Theoretical framework}
In this section the mathematical modeling of the proposed model is presented. We start with the necessary preliminaries concerning the incompatible elasticity theory of quasicrystals in subsection 3.1. For a more detailed exposition on this matter one can also see in \cite{Agiasofitouetal2010}. We continue in subsection 3.2 with the formulation of our problem based on a variational principle allowing for dissipation of the phason fields.
\subsection{Preliminaries}
The displacement field $\UU$ in a quasicrystal is constituted by two components, the usual {\it phonon displacement field} $u_i^\|(\Bx,t) \in E_\|$, which belongs to the physical or parallel space $E_\|$,
and the {\it phason displacement field} $u_i^\bot (\Bx, t) \in E_\bot$ that
belongs to the perpendicular space $E_\bot$, that is
\begin{equation}\label{U}
\UU=(\Bu^\|, \Bu^\bot) \ \in E_\| \oplus E_\bot.
\end{equation}
All quantities depend on the physical space coordinates $\Bx \in E_\|$ and the time~$t$. The differentiation of the displacement fields with respect to the physical coordinates gives the total distortion tensors for the phonon and phason fields, ${\beta}^{\|\, \T}_{ij}$ and ${\beta}^{\bot \T}_{ij}$, which can be decomposed into {\it the elastic distortion tensors} $\beta^\|_{ij}, \ \beta^\bot_{ij} $
and {\it the plastic distortion tensors} ${\beta}^{\|\, \P}_{ij}, \ {\beta}^{\bot \P}_{ij}$,
respectively,
\begin{alignat}{2}\label{ugrad}
u^\|_{i,j}:={\beta}^{\|\, \T}_{ij}=\beta^\|_{ij}+{\beta}^{\|\, \P}_{ij}, \qquad
u^\bot_{i,j}:={\beta}^{\bot \T}_{ij}=\beta^\bot_{ij}+{\beta}^{\bot \P}_{ij}.
\end{alignat}
A comma denotes differentiation with respect to the spatial physical coordinates. The differentiation of the displacement fields with respect to the time $t$ gives the total velocities, $v_i^{\|\, \T}$ and $v_i^{\bot\, \T}$, which can be decomposed into {\it the elastic velocities}, $v^\|_i$ and $v^\bot_i$, and {\it the plastic} or { \it initial velocities} \citep{Kossecka1975, Kossecka-deWit1977}, $v_i^{\|\, \P}$ and $v_i^{\bot\, \P}$, for the phonon and phason fields, respectively,
\begin{alignat}{2}\label{udot}
\dot{u}_i^\|:=v_i^{\|\, \T}=v_i^\|+v_i^{\|\, \P},\qquad
\dot{u}_i^\bot:=v_i^{\bot\, \T}=v_i^\bot+v_i^{\bot\, \P}.
\end{alignat}
A superimposed dot denotes differentiation with respect to time. The elastic distortion tensors and the elastic velocity vectors are the physical state quantities in the theory of quasicrystals.
\par
It should be emphasized that in the dislocation theory of
quasicrystals the incorporation of the plastic fields is essential, since a dislocation is the elementary carrier of plasticity. The plastic fields, especially the plastic distortion tensors, have to be included (see also~\citet{Agiasofitouetal2010}). For defects, the plastic distortion tensors give the so-called eigenstrains~\citep{Mura}. In the theory of linear elasticity of dislocations in quasicrystals the eigenstrains have been used by~\cite{Ding94, Ding95} and~\citet{Hu2000}. Another important aspect in the study of defects in quasicrystals that the reader should always take into consideration is that of the temperature. It is reported for example in \cite{Francoual2003} that phason fluctuations are being frozen at room temperature and plastic deformation
is possible through dislocations movements at enough high temperatures.

\subsection{Variational principal formulation with dissipation}
As we have discussed in Section 2, phasons should be modeled as damped waves. This means that dissipative processes are behind the damping. The presence of dissipative processes leads to the appearance of {\it phason frictional forces} (dissipative forces) which are linear functions of the phason velocities
\begin{align}\label{fr}
f_i^{\text{fr}}=-\D_{ij}v^\bot_j.
\end{align}
It is proven by methods of statistical physics \citep{Landau Statistical Physics} that the tensor $D_{ij}$ is symmetric, that is
\begin{align}\label{Dij}
\D_{ij}=\D_{ji}.
\end{align}
These forces can be written as the derivatives, with respect to the corresponding velocities, of the quadratic function
\begin{align}\label{R}
\RR=\frac{1}{2}\, \D_{ij}v^\bot_i v^\bot_j,
\end{align}
which is called {\it the dissipative function} \citep{ Landau Mechanics, Landau Statistical Physics, Landau Elasticity, Goldstein}.
Therefore,
\begin{align}\label{fr-R}
f_i^{\text{fr}}=-\frac{\partial \RR}{\partial v^\bot_i}.
\end{align}
The dissipative function gives the rate of dissipation of the energy in the system, making its physical significance very important.
It has been proven (see, e.g., \cite{Landau Mechanics, Landau Statistical Physics}) that $\RR$ should be positive definite ($\RR>0$), therefore the tensor $\D_{ij}$ should also be positive definite. We may call here $\D_{ij}$ the {\it tensor of phason friction coefficients}.

In the framework of incompatible elasticity theory, we consider
the Lagrangian density $\LL$ as a smooth function of the physical state quantities, that is $\LL=\LL(\Bv^\|, \Bv^\bot, \Bbeta^\|, \Bbeta^\bot)$ as well as the dissipation function density $\RR=\RR(\Bv^\bot)$ as it is given by the relation~(\ref{R}).  The corresponding Euler-Lagrange equations allowing for dissipative processes for the phason fields (\cite{Landau Mechanics, Landau Statistical Physics, Goldstein}) are
\begin{align}
&\label{EL1}\frac{\partial \LL}{\partial u^\|_i}-\frac{\partial}{\partial t}
\bigg(\frac{\partial \LL}{\partial \dot{u}^\|_i}\bigg)-
\frac{\partial}{\partial x_j}\bigg(\frac{\partial \LL}{\partial u^\|_{i,j}}\bigg)=0,\\
&\label{EL2} \frac{\partial \LL}{\partial u^\bot_i}
-\frac{\partial}{\partial t}\bigg(\frac{\partial \LL}{\partial \dot{u}^\bot_i}\bigg)-
\frac{\partial}{\partial x_j}\bigg(\frac{\partial \LL}{\partial u^\bot_{i,j}}\bigg)=\frac{\partial \RR}{\partial  \dot{u}^\bot_i}.
\end{align}
It can be seen that the Euler-Lagrange equation for the phason fields (eq.~(\ref{EL2})) differs from the usual form of the Euler-Lagrange equations by the presence of the derivative of the dissipative function density on the right-hand side. \\
Explicitly, the Lagrangian density is given in terms of the kinetic energy density $T$, the elastic energy density $W$ and the potential $V$ of the external forces
\begin{equation}
\LL=T-W-V.
\end{equation} The elastic energy density (for the unlocked state) is considered as a
quadratic function of the phonon and phason strains \citep{Ding1993, Agiasofitouetal2010}
\begin{equation}\label{w}
W=\frac{1}{2}\beta^\|_{ij}C_{ijkl}\beta^\|_{kl}+\beta^\|_{ij}D_{ijkl}\beta^\bot_{kl}+\frac{1}{2}\beta^\bot_{ij}E_{ijkl}\beta^\bot_{kl}
\end{equation}
with the tensors of elastic constants to possess the following symmetries
\begin{align}\label{Coef}
C_{ijkl}=C_{klij}=C_{ijlk}=C_{jikl},\quad D_{ijkl}=D_{jikl}, \quad E_{ijkl}=E_{klij}.
\end{align}
The kinetic energy density is considered to admit phonon and phason contributions
\begin{equation}\label{T3}
T=\frac{1}{2}\rho(v^\|_i)^2+\frac{1}{2}\rho(v^\bot_i)^2,
\end{equation}
where $\rho$ is the average mass density of the material.  \citet{Rochal2000} proposed that the phasons should possess
a generalized phason mass density, the  effective phason density  $\rho_{\text {eff}}$, which
may be different from the ordinary density $\rho$. However, later in their suggested minimal model \citep{Rochal2002}, they supported that the phason momentum in quasicrystals is not ``conserved", implying $\rho_{\text {eff}}=0$. In any case, the meaning of this quantity is not clarified and its measurement
is difficult (see, e.g., \cite{Fan Book}). In the case that the phasons indeed possess a different mass density, $\rho_{\text {eff}}$, the form of the kinetic energy density~(\ref{T3}) as well as of the corresponding derived equations bellow can be easily modified.\\
The potential $V$ can be expressed in terms of the {\it conventional (phonon) body force density} $f_i^\|$ and a  {\it generalized (phason) body force density} $f_i^\bot$ as follows
\begin{equation}\label{V3}
V=-u^\|_i f_i^\|-{u}_i^\bot f_i^\bot.
\end{equation}
The constitutive relations for the considered Lagrangian density are
\begin{align}
&p_i^\|=\frac{\partial T}{\partial v^{\|}_i}=\rho v^\|_i, \label{CR1}\\
&p^\bot_i=\frac{\partial T}{\partial v^\bot_i}=\rho v^\bot_i, \label{CR2}\\
&\sigma^\|_{ij}=\frac{\partial W}{\partial \beta^\|_{ij}}=C_{ijkl}\beta^\|_{kl}+D_{ijkl}\beta^\bot_{kl}, \label{CR3}\\
&\sigma^\bot_{ij}=\frac{\partial W}{\partial \beta^\bot_{ij}}=D_{klij}\beta^\|_{kl}+E_{ijkl}\beta^\bot_{kl} \label{CR4}.
\end{align}
In the above relations, $p_i^\|$ and $p^\bot_i$ are {\it the phonon} and {\it phason momentum vectors}
and $\sigma^\|_{ij}$ and $\sigma^\bot_{ij}$ are {\it the phonon} and {\it phason stress tensors}, respectively. We note that the phonon stress is symmetric,
$\sigma^\|_{ij}=\sigma^\|_{ji}$, but the phason stress
is asymmetric, $\sigma^\bot_{ij}\neq\sigma^\bot_{ji}$, (see also, e.g., \cite{Ding1993}).
\par

We should point out that the frictional force vector (\ref{fr}) is given as the scalar product of the anisotropic tensor $\D_{ij}$ and the phason velocity $\Bv^\bot$, in order to account for anisotropic effects of the damped waves. If we consider the case of an isotropic tensor $\D_{ij}$, that is $\D_{ij}=\D\delta_{ij}, \ \D>0$, then
\begin{align}\label{force}
f_i^{\text{fr}}=-\D {v}_i^\bot.
\end{align}
 $\D$ is the {\it phason friction coefficient}, $\D>0$. Such a force (eq.~(\ref{force}) in the compatible case) was introduced  under the name phason ``bulk force" for quasicrystals in order to capture an attenuation of diffusive type of the phason modes by \citet{Rochal2000, Rochal2002}. However, this force was actually used (\citet{Rochal2000, Rochal2002}) in the place of an external force and inserted ``by hand" in the derived equations of motion.  From the physical point of view, we notice the contradiction that a force depending on the velocity (internal field) has been considered as an external force. For that reason, a straightforward procedure, clear from the mathematical (derivation from Euler-Lagrange equations with dissipation) as well as from the physical point of view has been preferred and followed here.

\section{The equations of motion}

In this section, we first derive the field equations for the considered model in terms of the momenta and the stresses
as well as in terms of the elastic fields. Next, we give the equations of motion for the phonon and phason displacement fields in the theory of incompatible and compatible elastodynamics of wave-telegraph type of quasicrystals.
\par
The Euler-Lagrange equations~(\ref{EL1}) and (\ref{EL2}),
using the relations (\ref{ugrad}) and (\ref{udot}), the eq.~(\ref{fr-R}) as well as the
constitutive relations~(\ref{CR1})--(\ref{CR4}),
give {\it the field equations for the elastodynamic model of wave-telegraph
type of quasicrystals in terms of the momenta and the stresses}
\begin{align} \label{Eq.7}
\dot{p}_i^\|-\sigma^\|_{ij,j}&=f^\|_i,\\
\label{Eq.8}
\dot{p}_i^\bot-\sigma^\bot_{ij,j}-f_i^{\text{fr}}&=f^\bot_i.
\end{align}
The above equations, using the explicit formulas of the corresponding fields (see eqs. (\ref{CR1})--(\ref{CR4}) and eq.~(\ref{fr})),
can further give {\it the field equations in terms
of the elastic fields} $\Bv^\|, \Bv^\bot, \Bbeta^\|$ and $\Bbeta^\bot$ as follows
\begin{align}\label{Eq.9}
\rho \dot{v}^\|_i-C_{ijkl}\beta^\|_{kl,j}-D_{ijkl}\beta^\bot_{kl,j}&=f^\|_i, \\
\label{Eq.10}
\rho \dot{v}^\bot_i+\D_{ij}{v}_j^\bot-
D_{klij}\beta^\|_{kl,j}-E_{ijkl}\beta^\bot_{kl,j}&=f^\bot_i.
\end{align}
Eqs.~(\ref{Eq.9}) and (\ref{Eq.10}), using eqs.~(\ref{ugrad}) and
(\ref{udot}), give {\it the equations of motion for the phonon and phason displacement fields}
\begin{align}
\label{Eq.11}
\rho \ddot{u}_i^\|-C_{ijkl}u_{k,lj}^\|-D_{ijkl}u_{k,lj}^\bot&=
\rho \dot{v}_i^{\|\, \P}-C_{ijkl}{\beta}^{\|\, \P}_{kl,j}-D_{ijkl}{\beta}^{\bot\, \P}_{kl,j}+f_i^\|,\\
\label{Eq.12}
\rho \ddot{u}_i^\bot+\D_{ij}\dot{u}_j^\bot
-D_{klij}u_{k,lj}^\|-E_{ijkl}u_{k,lj}^\bot&=
\rho \dot{v}_i^{\bot\, \P}
+\D_{ij}{v}_j^{\bot\, \P}
-D_{klij}{\beta}^{\|\, \P}_{kl,j}-E_{ijkl}{\beta}^{\bot\, \P}_{kl,j}+f^\bot_i,
\end{align}
where the plastic fields and the phonon and phason body forces are
the source terms.
Eq.~(\ref{Eq.11}) is a partial differential equation of wave
type, or in other words an elastodynamic Navier equation and
eq.~(\ref{Eq.12}) is a partial differential equation of telegraph type. Both equations are hyperbolic. The system of the partial differential equations~(\ref{Eq.11}) and (\ref{Eq.12}) is coupled due
to the tensor of phonon-phason coupling $D_{ijkl}$. Special attention should be focused on the term $\D_{ij} \dot{u}_j^\bot$ in eq.~(\ref{Eq.12}), which contains the first time derivative of the phason displacement field. This term is responsible for the damping and appears due to the existence of the phason frictional forces. To sum up, {\it eqs.~(\ref{Eq.11}) and (\ref{Eq.12}) are the equations of motion for the displacement fields
in the theory of incompatible elastodynamics of wave-telegraph type of quasicrystals}.

\par
In the theory of dislocations \citep{Lardner}, the concept of the plastic velocity, introduced by \citet{Kossecka1969} (see also \citet{Kossecka1975, Kossecka-deWit1977}), has not been widely recognized and used. For that reason, we examine also the ``usual" case of elastoplasticity, which is obtained when
the plastic velocities are zero, that is, $v_i^{\|\, \P}=0,\quad v_i^{\bot\, \P}=0$.
Then, eqs.~(\ref{Eq.11}) and (\ref{Eq.12}) (the external forces are zero) reduce to the following equations of motion
\begin{align}
\label{Eq.11-v}
&\rho \ddot{u}_i^\|-C_{ijkl}u_{k,lj}^\|-D_{ijkl}u_{k,lj}^\bot=
-C_{ijkl}{\beta}^{\|\, \P}_{kl,j}-D_{ijkl}{\beta}^{\bot\, \P}_{kl,j},\\
\label{Eq.12-v}
&\rho \ddot{u}_i^\bot+\D_{ij}\dot{u}_j^\bot
-D_{klij}u_{k,lj}^\|-E_{ijkl}u_{k,lj}^\bot=
-D_{klij}{\beta}^{\|\, \P}_{kl,j}-E_{ijkl}{\beta}^{\bot\, \P}_{kl,j}.
\end{align}
In this case, only the plastic distortion tensors,
$\beta^{\|\, \P}_{ij}$ and $\beta^{\bot \P}_{ij}$, are non-vanishing and play
the physical role of the sources for the dislocation displacement fields, $u_i^\|$ and $u_i^\bot$.
Thus, it becomes evident that the plastic distortion tensors cannot be neglected in the study of dislocations.

\par
We now proceed to investigate the compatible case. In the compatible elastodynamics all the plastic fields are zero, that is, $\beta^{\|\, \P}_{ij}=0$, $\beta^{\bot \P}_{ij}=0$, $v_i^{\|\, \P}=0$ and $v_i^{\bot\, \P}=0$.
In this case, eqs.~(\ref{Eq.7}) and (\ref{Eq.8}), and eqs.~(\ref{Eq.9}) and (\ref{Eq.10}), are valid and eqs.~(\ref{Eq.11}) and (\ref{Eq.12}) are reduced to the following ones
\begin{align}
\label{Eq.13}
&\rho \ddot{u}_i^\|-C_{ijkl}u_{k,lj}^\|-D_{ijkl}u_{k,lj}^\bot=f_i^\|,\\
\label{Eq.14}
&\rho \ddot{u}_i^\bot+\D_{ij}\dot{u}_j^\bot
-D_{klij}u_{k,lj}^\|-E_{ijkl}u_{k,lj}^\bot=f^\bot_i.
\end{align}
The mathematical character of the above equations remains the same as of the corresponding ones
in the incompatible case, that is, eq.~(\ref{Eq.13}) is a partial differential equation of wave type
and  eq.~(\ref{Eq.14}) is of telegraph type. {\it Eqs.~(\ref{Eq.13}) and (\ref{Eq.14}) are  the equations of motion for the displacement fields
in the theory of compatible elastodynamics of wave-telegraph type of quasicrystals}.
\par
The tensor telegraph differential operator (see relation~(\ref{L-T})) is the same for all three equations~(\ref{Eq.12}), (\ref{Eq.12-v}), and (\ref{Eq.14}). Comparing the scalar telegraph operator in eq.~(\ref{tel}) with the tensor telegraph operator in relation~(\ref{L-T}), one may define {\it the tensor of characteristic time of damping}, $ \tau_{ij}$, as follows
\begin{align}\label{t-ij-tel}
   ( \tau_{ij})_{\text {tel}}=2 \rho \D_{ij}^{-1},
\end{align}
where $\D_{ij}^{-1}$ is the inverse tensor of the $\D_{ij}$. It is reasonable that the tensor of characteristic time of damping depends on the average mass density of the material and the tensor of phason friction coefficients. Particularly, we see that the anisotropy of the tensor $\D_{ij}$ influences the tensor of characteristic time of damping. It is obvious that $(\tau_{ij})_{\text{tel}}=(\tau_{ji})_{\text {tel}}$.
\par
In the isotropic case of the phason friction tensor $\D_{ij}$, that is  $\D_{ij}=\D\delta_{ij}$, {\it the characteristic time of damping} simplifies to
\begin{align}\label{t-tel}
\tau_{\text {tel}}=\frac{2 \rho}{\D},
\end{align}
and the corresponding eqs.~(\ref{Eq.10}), (\ref{Eq.12}), (\ref{Eq.12-v}), and (\ref{Eq.14}) can be easily modified.  From eq.~(\ref{t-tel}), it is clear that\footnote{The phason friction coefficient $\D$  should not be confused with the ``phason diffusion constant", $D_{phason}=1/{\tau q^2}$ \citep{de Boissieu et al 2008}. It is apparent that the two quantities have different units. $D_{phason}$ is the $\D_{\text{dif}}$ in our notation (see eq.~(\ref{theta})).}  the dimension of $\D$ is $[\D]=\frac{[M]}{[L]^3[T]}$ and coincides with the dimension of  ${1}/{\Gamma_w}$, where $\Gamma_w$ is the so-called phason kinetic coefficient introduced by \citep{Lubensky85, Lubensky86}. Hence, $[\D]=1/[\Gamma_w]$.

\par
 Making a qualitative asymptotic analysis to the telegraph type equation (see, e.g., eq.~(\ref{Eq.14})) one can see that if the friction parameter $\D_{ij}$ is very large compared with the average mass density $\rho$ per unit time, then the second term of the telegraph equation is dominant in comparison with the first term. In this case, the telegraph type equation degenerates to a diffusion type equation\footnote{In this case, the phason diffusion parameter can be interrelated with the phason friction coefficient: $[\D_{\text {dif}}]\sim[E_{ijkl}]/[\D_{ij}]$.}. If $\D_{ij}$ is very small, then the wave character is dominant. For intermediate values of the $\D_{ij}$, we have waves with attenuation. This qualitative analysis of $\D_{ij}$ shows how the proposed model constitutes a unified theory of the previously mentioned models.

\par
 It is important to notice that the characteristic time of damping $\tau_{\text{tel}}$ given by eq.~(\ref{t-tel}) is qualitatively and quantitatively different from the relaxation time (eq.~(\ref{theta})) derived from the dispersion relation of the diffusion equation \citep{Francoual2003, de Boissieu et al 2008}. The characteristic time of damping $\tau_{\text{tel}}$ is appearing in the equation of motion itself and it is not based on dispersion relations. It is remarkable that for sufficiently large wavevectors it coincides with the relaxation time $\tau$ (see eq.~(\ref{t=tel})) produced by the dispersion relation of the telegraph equation as explained in the Appendix \ref{appendixA}.
\par
  A general view into the proposed model concerning the equation of motion of the phason displacement field shows that the
consideration of two quantities is important; that of the phason momentum vector and of the phason frictional forces. If the phason momentum
is zero, then eq.~(\ref{Eq.12}) (or eq.~(\ref{Eq.12-v}), eq.~(\ref{Eq.14})) will be a partial differential equation
of diffusion type (infinite velocity of propagation) and if the phason frictional forces are zero, then eq.~(\ref{Eq.12}) (or eq.~(\ref{Eq.12-v}), eq.~(\ref{Eq.14})) will be a partial differential equation
of wave type (without damping). Hence, it is evident that we need both quantities in order to obtain the desired equation
for the phason displacement field, that is the equation of telegraph type.

\section{Conclusions}
We have proposed the elastodynamic model of wave-telegraph type to describe the dynamics of quasicrystals.
The main feature of this model is that the phason modes are described by a telegraph type partial differential
equation and consequently represent waves damped in time with finite propagation velocity. The telegraph equation has drastically different solutions compared with the solutions provided by the diffusion equation. Since the two equations of motion are coupled, the choice of a new equation for the phason fields will have a very different impulse to phonon dynamics also. Therefore, the dynamic behavior of the whole system is severely influenced by the adoption of a different equation for the description of phason dynamics.
\par
Several noteworthy features characterize the proposed model. By construction, a telegraph type equation supports wave propagation with attenuated amplitude. The influence of the damping in the dynamic behavior of the phasons is expressed by the tensor of phason friction coefficients, which gives the possibility to take into account that the phason waves can be damped anisotropically. In terms of the phason friction coefficient and the average mass density of the material an important quantity, the characteristic time of damping, has been defined. Moreover, this model provides a theory valid in the whole regime of possible wavelengths for the phasons. In addition, with the new model there is no longer the drawback of the infinite propagation velocity that exists with the equation of diffusion type.
\par
 One more advantage of a telegraph type equation is that it establishes uniformity and symmetry between the phonons and phasons as far as the corresponding initial value problems are concerned. More precisely, the implication of diffusion equation creates a lack of symmetry to the whole system. Indeed, the phonons are represented by a wave type partial differential equation which leads to a solution depending on two initial data, that is,  the initial values of both phonon displacement and total velocity fields. If the phasons are represented by a diffusion type equation, then the solution depends only on the initial condition for the displacement field (and the initial total velocity is absent). While if phasons are represented by a telegraph type equation, the solution incorporates both initial data for phason displacement and total velocity fields. Therefore, a telegraph type equation restores uniformity in the settlement of initial value problems.
\par
The understanding of the phason modes is not only important for the study of quasicrystals but also of other types of aperiodic crystals. It seems that the
proposed model could describe very well the behavior of incommensurately modulated phases. Indeed, it has been observed by \citet{deBoissieu2008} that there are two regimes for the phason modes, a long-wavelength regime where the phasons are diffusive and a shorter
wavelength regime where the phasons are damped propagating modes. Therefore, the proposed model could give a unified and broad framework for the description of aperiodic crystals. Of course, its establishment depends on its
experimental validation. We believe that the introduction of the new model will bring a new perspective, promoting the
study in the field of quasicrystals and further of aperiodic crystals.

\section*{Acknowledgements} 
Dr. Eleni Agiasofitou is greatly indebted to Professor Antonios Charalambopoulos 
for efficient discussions and valuable remarks concerning several aspects 
of wave propagation. 
The authors gratefully acknowledge grants from the Deutsche Forschungsgemeinschaft 
(Grant Nos. La1974/2-1, La1974/2-2, La1974/3-1).

\begin{appendix}

\section{Dispersion relations}
\label{appendixA}
\setcounter{equation}{0}
\renewcommand{\theequation}{\thesection.\arabic{equation}}

  The dispersion relations contain a lot of information about the behavior of wave propagation inside a medium. Here, we give the dispersion relations for the diffusion equation (we examine the case of standing waves damped exponentially in time for necessary comparisons) and a typical telegraph equation. For sake of simplicity, this appendix is restricted to one space dimension.
 \par
 A plane harmonic wave moving in the $z$-direction is given by
\begin{align}\label{wave}
\psi(z,t)=A\, \e^{\ii (qz-\omega t)},
\end{align}
where $A$ is the wave amplitude, $q={2\pi}/{\lambda}$ stands for the wavenumber,  $\lambda$ is the corresponding wavelength, and $\omega$ is the circular frequency.
\par
The diffusion equation is given by
\begin{align}\label{Dif}
\frac{\partial \psi}{\partial t}=\D_{\text {dif}}\, \frac{\partial^2 \psi}{\partial z^2}, \quad \D_{\text {dif}}>0.
\end{align}
Application of eq.~(\ref{wave}) to eq.~(\ref{Dif}) yields to the dispersion relation \citep{Kneubuhl}
\begin{align}\label{disp-dif}
\omega=-\ii\, \D_{\text{dif}}\, q^2.
\end{align}
Substituting eq.~(\ref{disp-dif}) into eq.~(\ref{wave}) leads to damped standing waves represented by
\begin{align}\label{standing}
\psi(z,t)=A\, \e^{\ii qz} \e^{-\theta t}=A\, \e^{\ii qz} \e^{-{t}/{\tau}},
\end{align}
 where the damping coefficient $\theta$ is given by
\begin{align}\label{theta}
\theta=\frac{1}{\tau}= \D_{\text {dif}}\, q^2
\end{align}
with $\tau$ the corresponding relaxation time.
\par
Let us consider now the telegraph equation (\ref{tel}) in one space dimension
 \begin{align}    \label{tele}
 \frac{\partial^2 \psi}{\partial t^2} +\frac{2}{\tau_ {\text {tel}}}\, \frac{\partial\psi}{\partial t}=c^2 \frac{\partial^2 \psi}{\partial z^2}.
 \end{align}
Substituting eq.~(\ref{wave}) into eq.~(\ref{tele}), we find that $\omega$ and $q$ must satisfy the dispersion relation \citep{Jeffrey2003}
\begin{align}\label{om}
\omega^2+\frac{2\ii}{\tau_ {{\text {tel}}}}\, \omega-c^2 q^2=0.
\end{align}
We can distinguish three cases:\\
i) $c\,q > {1}/\tau_ {\text {tel}}$\\
Then, from eq.~(\ref{om}) we deduce
\begin{align}\label{omega}
\omega=-\frac{\ii}{\tau_ {\text {tel}}} \pm \sqrt{c^2\, q^2\,-\frac{1}{\tau_ {\text {tel}}^2}}
\end{align}
and consequently the solution (\ref{wave}) takes the form
\begin{align}\label{finalpsi}
\psi(z,t)=A\, \e^{-\frac{1}{\tau_ {\text {tel}}}t}\, \e^{\ii \bigl(qz \mp \sqrt{c^2q^2-
\frac{1}{\tau^2_ {\text {tel}}}}\, t \bigl)}.
\end{align}
 Eq.~(\ref{finalpsi}) represents a traveling wave damped in time with relaxation time $\tau$, which is equal to the  characteristic time of damping $\tau_ {\text {tel}}$ \begin{align}\label{t=tel}
\tau=\tau_ {\text {tel}}
\end{align}
and phase velocity
\begin{align}\label{c_p}
c_p=c\, \sqrt{1-\frac{1}{c^2\, \tau^2_{\text {tel}}\, q^2}}.
\end{align}
ii) $c\,q < {1}/\tau_ {\text {tel}}$\\
Then, we deduce from eq.~(\ref{wave}) that only standing waves damped in time are supported of the form
\begin{align}
\psi(z,t)=A\, \e^{\ii qz} \e^{-\theta t}=A\, \e^{\ii qz} \e^{-{t}/{\tau}}
\end{align}
where two relaxation times appear
\begin{align}\label{eq12}
\theta=\frac{1}{\tau}=\frac{1}{\tau_{{\text {tel}}}}\pm\sqrt{\frac{1}{\tau^2_{\text {tel}}}-c^2\,q^2}.
\end{align}
 $\theta >0$ for both values.\\
 iii) $c\,q = {1}/\tau_ {\text {tel}}$\\
In this case, it is easy to see that we have standing waves damped in time with relaxation time equal to the characteristic time of damping, $\tau=\tau_ {\text {tel}}$.
\par
Consequently, the critical value $q_0={1}/{c\, \tau_ {\text {tel}}}$ can be considered as a crucial threshold for the wavenumber $q$. If $q<q_0$, then we have only exponentially damped standing waves with two relaxation times. When $q$ overpasses $q_0$, then propagating waves (damped in time) emerge with relaxation time equals the characteristic time of damping and phase velocity dependent on the wavenumber $q$ via the relation (\ref{c_p}). If $q=q_0$, then standing waves appear with relaxation time equal to the characteristic time of damping.
\par
 All these arguments can be reformulated in terms of the wavelength $\lambda$ of the process. In case that $\lambda > \lambda_0$, where $\lambda_0=2\pi c\,\tau_{\text{tel}}$ is the critical wavelength, then standing waves appear with two characteristic times depending on $q$  via eq.~(\ref{eq12}) which are different from the relaxation time (\ref{theta}). In case that $\lambda < \lambda_0$, then propagating waves damped in time of the form (\ref{finalpsi}) emerge. This has a physical interest in case that  $\lambda_0$ is sufficiently larger than the characteristic dimension $d_0$ of the interatomic distance, so that the wavelength $\lambda$ satisfies the basic principle of the continuum theory to remain larger than $d_0$.

\section{Matrix form of the coupled equations of motion of the elastodynamic model of wave-telegraph type}
\label{appendixB}
\setcounter{equation}{0}
\renewcommand{\theequation}{\thesection.\arabic{equation}}

For a numerical implementation it is useful to rewrite the equations of motion in a matrix
form. First, we define the following tensor differential operators:
\begin{itemize}
\item
tensor wave operator:\\
\begin{align}
\label{L-W}
L_{ik}^{\text{w}}=\rho\, \delta_{ik}\pd_{tt}-C_{ijkl}\pd_j\pd_l
\end{align}
\item
tensor telegraph operator:\\
\begin{align}
\label{L-T}
L_{ik}^{\text{te}}=\rho\, \delta_{ik}\pd_{tt}+\D_{ik}\, \pd_t -E_{ijkl}\pd_j\pd_l
\end{align}
\item
tensor operator of phonon-phason coupling:\\
\begin{align}
\label{L-C}
L_{ik}^{\text{c}}=-D_{ijkl}\, \pd_j \pd_l,
\end{align}
\end{itemize}
where $\pd_t$ and $\pd_j$ are the first derivatives with respect to the time and the spatial coordinates $x_j$, respectively.
\par
Using the tensor differential operators~(\ref{L-W})--(\ref{L-C}),
the coupled equations of motion (\ref{Eq.13}) and (\ref{Eq.14}) for the compatible case
read in the matrix form
\begin{align}
\label{MR}
\begin{pmatrix}
L_{ik}^{\text{w}} & L_{ik}^{\text{c}}\\
L_{ki}^{\text{c}} & L_{ik}^{\text{te}}
\end{pmatrix}
\begin{pmatrix}
u_{k}^{\|}\\
u_{k}^{\bot}
\end{pmatrix}
=\begin{pmatrix}
f_{i}^{\|}\\
f_{i}^{\bot}
\end{pmatrix},
\end{align}
and the coupled equations of motion~(\ref{Eq.11}) and (\ref{Eq.12}) for the incompatible case are given by
\begin{align}
\label{MR2}
\begin{pmatrix}
L_{ik}^{\text{w}} & L_{ik}^{\text{c}}\\
L_{ki}^{\text{c}} & L_{ik}^{\text{te}}
\end{pmatrix}
\begin{pmatrix}
u_{k}^\|\\
u_{k}^\bot
\end{pmatrix}=
\begin{pmatrix}
\rho\,\delta_{ik} \pd_t & 0\\
 0 &\rho\, \delta_{ik}\pd_t+\D_{ik}
\end{pmatrix}
\begin{pmatrix}
v_{k}^{\|\,\P }\\
v_{k}^{\bot\,\P }
\end{pmatrix}
-\begin{pmatrix}
C_{ijkl}\pd_j & D_{ijkl}\pd_j \\
D_{klij}\pd_j  &E_{ijkl}\pd_j
\end{pmatrix}
\begin{pmatrix}
\beta_{kl}^{\|\,\P }\\
\beta_{kl}^{\bot\,\P }
\end{pmatrix}
+
\begin{pmatrix}
f_{i}^{\|}\\
f_{i}^{\bot}
\end{pmatrix}\,.
\end{align}

\end{appendix}

\end{document}